\documentclass[11pt]{article}%
\usepackage{graphicx}
\usepackage{amsmath}
\usepackage{amsfonts}
\usepackage{amssymb}
\usepackage{graphicx}%
\providecommand{\U}[1]{\protect\rule{.1in}{.1in}}
\begin{document}

\begin{center}

\bigskip

{\Large \textbf{Quasi-distributions for arbitrary non-commuting operators}}

\bigskip

J.\ S.\ Ben-Benjamin\textsuperscript{1}, L.\ Cohen\textsuperscript{2}
\end{center}

\textsuperscript{1} \textit{Institute for Quantum Science and Engineering,
Texas A\&M University, Texas, USA}

\textsuperscript{2} \textit{Hunter College and Graduate Center, City
University of New York, NY, USA}


\section*{Abstract}

We present a new approach for obtaining quantum quasi-probability
distributions, $P(\alpha,\beta),$ for two arbitrary operators, $\mathbf{a}$
and $\mathbf{b,}$ where $\alpha$ and $\beta$ are the corresponding
c-variables. We show that the quantum expectation value of an arbitrary
operator can always be expressed as a phase space integral over $\alpha$ and
$\beta,$ where the integrand is a product of two terms: One dependent only on
the quantum state, and the other only on the operator. In this formulation,
the concepts of quasi-probability and correspondence rule arise naturally in
that simultaneously with the derivation of the quasi-distribution, one obtains
the generalization of the concept of correspondence rule for arbitrary operators.

\section{Introduction}

We present a new approach for obtaining quasi-probability distributions,
$P(\alpha,\beta),$ for two arbitrary operators, $\mathbf{a}$ and $\mathbf{b,}$
where $\alpha$ and $\beta$ are the corresponding c-variables. The method also
yields the correspondence rule relating a quantum operator, $\mathbf{g,}$ to
the corresponding c-function, $g(\alpha,\beta)$. We assume total generality
for the operators.

Historically, quasi-probability distributions have been studied for the
position-momentum case ($qp$ case). Parallel to this development was the
development aimed at finding the quantum operator for a given classical
function. Such rules are called correspondence rules or rules of association.
Moyal \cite{moyal} showed that the Wigner distribution \cite{wigner} can be
derived using the Weyl correspondence rule \cite{weyl} and Cohen showed
\cite{cohen66} how to derive an infinite number of quasi-distributions and the
associated correspondence rules. The methods used in those studies have been
by way of the characteristic function operator. Scully \cite{mos,ben1,ben2}
emphasized a different approach for obtaining quasi-distributions of position
and momentum, namely to \textit{start} with quantum mechanics; One starts with
the quantum expectation value of an arbitrary operator, $\left\langle
\mathbf{g}\right\rangle =$Tr$[$$\boldsymbol{\mathbf{g}}$$\boldsymbol{\rho}],$
where $\mathbf{g}$ is the operator and $\boldsymbol{\rho}$ is the density
matrix, and shows that Tr$[$$\boldsymbol{\mathbf{g}}$$\boldsymbol{\rho}]$ can
always be expressed as a phase space integral over position and momentum.
Crucially the integrand factorizes into two terms: One dependent only on the
density matrix, and the other only on the quantum operator. In this way, one
not only derives the quasi-probability distributions, but also what have been
called the inverse correspondence rules \cite{kim,cohen-book-weyl}. In this
formulation, the concepts of quasi-probability and correspondence arise
naturally and together.

To illustrate the basic idea, we give the original derivation of Scully. The
expectation value of an operator $\mathbf{g}$ is\footnote{We often represent
multiple integrals by a single integration symbol, the differentials
indicating the number of integrals. All integrals go from $-\infty$ to
$\infty$.}
\begin{equation}
\langle\mathbf{g}\rangle=\text{Tr}[\boldsymbol{\mathbf{g}\rho}]=\int\,\langle
q|\boldsymbol{\mathbf{g}\rho}|q\rangle dq \label{step1}%
\end{equation}
Inserting the identity operator three times, we have
\begin{equation}
\langle\mathbf{g}\rangle=\iiiint\,\langle p_{1}|\mathbf{g}|p_{2}\rangle\langle
p_{2}|q_{1}\rangle\langle q_{1}|\boldsymbol{\rho}|q_{2}\rangle\langle
q_{2}|p_{1}\rangle dq_{1}dq_{2}dp_{1}dp_{2} \label{step2}%
\end{equation}
where bras and kets of $p_{1}$ and $p_{2}$ are momentum eigenstates, and those
of $q_{1}$ and $q_{2}$ correspond to position eigenstates. We change variables
in Eq. \eqref{step2}, defining the difference between the positions and
between the momenta as $\bar{q}=q_{1}-q_{2}$ and $\bar{p}=p_{2}-p_{1}$, and
the average position and momentum as $q=\frac{1}{2}(q_{1}+q_{2})$ and
$p=\frac{1}{2}(p_{1}+p_{2})$, respectively. Eq. \eqref{step2} then becomes
\begin{equation}
\langle\mathbf{g}\rangle=\iint\left(  \int\,\left\langle p-\frac{1}{2}\bar
{p}\middle|\mathbf{g}\middle|p+\frac{1}{2}\bar{p}\right\rangle e^{-i\bar
{p}q/\hbar}d\bar{p}\right)  \left(  \int\,\left\langle q+\frac{1}{2}\bar
{q}\left\vert \frac{\boldsymbol{\rho}}{2\pi\hbar}\right\vert q-\frac{1}{2}%
\bar{q}\right\rangle e^{-i\bar{q}p/\hbar}d\bar{q}\right)  dqdp\, \label{step3}%
\end{equation}
which shows that the quantum expectation value can be written as a phase space
integral; moreover, it shows that the integrand has been factorized as the
product of two c-functions. The first factor is the classical function,
$g(q,p),$ that corresponds to the operator $\mathbf{g}$ by way of the
inverse-Weyl rule \cite{kim} and the second factor depends only on the density
matrix $\boldsymbol{\rho}$, which in this case is the Wigner distribution.
Recently the above program has been carried out for the general class of
quasi-distributions of position and momentum \cite{ben2}.

In this paper we generalize the case of position and momentum to arbitrary
operators. We start with the expectation value of the operator $\mathbf{g,}$
as per Eq. \eqref{step1}
\begin{equation}
\left\langle \mathbf{g}\right\rangle =\text{Tr}[\boldsymbol{\mathbf{g}\rho}]
\label{eq:67-3}%
\end{equation}
and show that this can always be written in the form
\begin{equation}
\left\langle \mathbf{g}\right\rangle =\iint\left\{
\begin{array}
[c]{c}%
\text{a c-function of }\alpha\text{ and }\beta\text{ }\\
\text{that depends \textit{only} on }\mathbf{g}%
\end{array}
\right\}  \left\{
\begin{array}
[c]{c}%
\text{a c-function of }\alpha\text{ and }\beta\text{ }\\
\text{that depends \textit{only} on }\boldsymbol{\rho}%
\end{array}
\right\}  d\alpha d\beta\label{eq:67-4}%
\end{equation}
The left hand factor is the c-function, $g(\alpha,\beta),$ corresponding to
the operator and the right hand factor of the integrand is the
quasi-distribution, $P(\alpha,\beta).$ What Eq.\ \eqref{eq:67-4} shows is that
we can calculate the expectation value of a quantum operator by way of phase
space averaging in the phase space of $\alpha$ and $\beta$,
\begin{equation}
\left\langle g(\alpha,\beta)\right\rangle =\iint g(\alpha,\beta)P(\alpha
,\beta)d\alpha d\beta
\end{equation}
and that it equals $\left\langle \mathbf{g}\right\rangle $ calculated quantum
mechanically as per Eq.\ \eqref{eq:67-3}.

There has been some previous work on obtaining quasi-probability distributions
for quantities other than position and momentum. In particular, the case of
spin components \cite{spin1,chan}, local energy and local kinetic energy
\cite{bader1,bader2,ke,muga3,mazz}, and local spread of an operator (the
conditional standard deviation) have been considered
\cite{muga4,sala,loug-curr}. {Scully and M. S. Zubairy }\cite{sz} and
{Schleich }\cite{sc} used a similar approach to the one presented here to
derive the Q distribution. Also, similar considerations arise in the field of
time-frequency analysis \cite{bara,cohen-rev, tf-book,ins}. Previous attempts
to generalize to arbitrary operators have been approached by way of the
characteristic function operator method, but the general difficulty with that
method is the disentanglement of the exponential
\cite{scully-cohen,arb,cohenieee,cohenieee2}.

We present our results by first deriving two special cases and then deriving a
general form that generates an infinite number of distributions characterized
by a kernel function. We deal with the continuous case, and the results for
the discrete case follow naturally.

\section{Notation, marginals, density matrix, and transformation matrix}

The measurable quantities of the operators $\mathbf{a}$ and $\mathbf{b}$ are
denoted by $\alpha$ and $\beta$, which are the eigenvalues of the respective
operators obtained by solving the eigenvalue problem for each operator {%
\begin{equation}
\mathbf{a}u_{\alpha}(q)=\alpha u_{\alpha}(q)\hspace{0.5in};\hspace
{0.5in}\mathbf{b}v_{\beta}(q)=\beta v_{\beta}(q)
\end{equation}
where }$u_{\alpha}(q)$ and $v_{\beta}(q)$ are the corresponding eigenfunctions
in the position representation. The{ quantum probabilities for }$\alpha$ and
$\beta$ { are respectively }%
\begin{equation}
P(\alpha)=\left\vert \mathcal{A}(\alpha)\right\vert ^{2}=\left\vert \int
\psi(q)u_{\alpha}^{\ast}(q)\ dq\right\vert ^{2}\hspace{0.5in};\hspace
{0.5in}P(\beta)=\left\vert \mathcal{B}(\beta)\right\vert ^{2}=\left\vert
\int\psi(q)v_{\beta}^{\ast}(q)\ dq\right\vert ^{2}%
\end{equation}
where the wave function is expanded in terms of the eigenfunctions of
$\mathbf{a}$ or of $\mathbf{b}$
\begin{equation}
\psi(q)=\int\mathcal{A}(\alpha)u_{\alpha}(q)\ d\alpha\hspace{0.5in}%
;\hspace{0.5in}\psi(q)=\int\mathcal{B}(\beta)v_{\beta}(q)\ d\beta
\end{equation}
with $\mathcal{A}(\alpha)$ and $\mathcal{B}(\beta)$, respectively,
\begin{equation}
\mathcal{A}(\alpha)=\int\psi(q)u_{\alpha}^{\ast}(q)dq\hspace{0.5in}%
;\hspace{0.5in}\mathcal{B}(\beta)=\int\psi(q)v_{\beta}^{\ast}(q)dq
\end{equation}
In wave function terminology, $\mathcal{A}(\alpha)$ is the wave function in
the $\alpha$ representation and $\mathcal{B}(\beta)$ is the wave function in
the $\beta$ representation \cite{bohm}. { }

For the quasi-distribution, $P(\alpha,\beta),$ the marginal conditions
are$,\,$%
\begin{align}
\int P(\alpha,\beta)d\beta &  =\left\vert \mathcal{A}(\alpha)\right\vert
^{2}=\left\vert \int\psi(q)u_{\alpha}^{\ast}(q)\ dq\right\vert ^{2}\\
\int P(\alpha,\beta)d\alpha &  =\left\vert \mathcal{B}(\beta)\right\vert
^{2}=\left\vert \int\psi(q)v_{\beta}^{\ast}(q)\ dq\right\vert ^{2}%
\end{align}

\noindent\textbf{Density matrix. }In the $\alpha$ representation, the pure
state density matrix is
\begin{equation}
\rho(\alpha^{\prime\prime},\alpha^{\prime})=\mathcal{A}^{\ast}(\alpha^{\prime
})\mathcal{A}(\alpha^{\prime\prime})
\end{equation}
and the matrix elements $g_{\alpha^{\prime}\alpha^{\prime\prime}}$ of the
operator $\mathbf{g(x,p}_{x}\mathbf{)}$ in the $\alpha$ basis are
\begin{equation}
g_{\alpha^{\prime}\alpha^{\prime\prime}}\mathbf{=}\int\delta(\alpha^{\prime
}-x)\mathbf{g(x,p}_{x}\mathbf{)}\delta(\alpha^{\prime\prime}-x)dx
\end{equation}
In general, the expectation value of $\mathbf{g}$ is
\begin{equation}
\left\langle \mathbf{g}\right\rangle =\iint\rho(\alpha^{\prime\prime}%
,\alpha^{\prime})g_{\alpha^{\prime}\alpha^{\prime\prime}}\;d\alpha^{\prime
}d\alpha^{\prime\prime}%
\end{equation}
Explicitly,
\begin{equation}
\left\langle \mathbf{g}\right\rangle =%
{\displaystyle\iiint}
\mathcal{A}^{\ast}(\alpha^{\prime})\mathcal{A}(\alpha^{\prime\prime}%
)\delta(\alpha^{\prime}-x)\mathbf{g}\delta(\alpha^{\prime\prime}%
-x)d\alpha^{\prime}d\alpha^{\prime\prime}dx\label{eq:01-15}%
\end{equation}
Eq.\ \eqref{eq:01-15} will be our starting point.

\bigskip

\noindent\textbf{Transformation matrix. }We define the transformation matrix,
$T(\beta,\alpha),$ from the $\alpha$ to the $\beta$ representation by
\cite{bohm}
\begin{equation}
T(\beta,\alpha)=\int v_{\beta}^{\ast}(q)u_{\alpha}(q)\,dq
\end{equation}
in which case, the eigenfunctions are related by
\begin{align}
u_{\alpha}(q) &  =\int T(\beta,\alpha)v_{\beta}(q)\,d\beta\\
v_{\beta}(q) &  =\int T^{\ast}(\beta,\alpha)u_{\alpha}(q)d\alpha=\int
T^{\dagger}(\alpha,\beta)u_{\alpha}(q)d\alpha
\end{align}
and the wave functions $\mathcal{A}(\alpha)$ and $\mathcal{B}(\beta)$ are
related by
\begin{align}
\mathcal{A}(\alpha) &  =\int\mathcal{B}(\beta)T^{\ast}(\beta,\alpha)d\beta\\
\mathcal{B}(\beta) &  =\int\mathcal{A}(\alpha)T(\beta,\alpha)d\alpha
\end{align}
The transformation matrix satisfies
\begin{equation}
\int T^{\dagger}(\alpha,\beta)T(\beta,\alpha^{\prime})d\beta=\delta
(\alpha-\alpha^{\prime})\label{eq19}%
\end{equation}
and \
\begin{equation}
T^{\dagger}(\beta,\alpha)=T^{\ast}(\alpha,\beta)
\end{equation}

\section{Wigner type case}

We start with the quantum expectation value, Eq.\ \eqref{eq:01-15}
\begin{equation}
\left\langle \mathbf{g}\right\rangle =\int\mathcal{A}^{\ast}(\alpha^{\prime
})\mathcal{A}(\alpha^{\prime\prime})\ \delta(\alpha^{\prime}-x)\mathbf{g}%
\delta(\alpha^{\prime\prime}-x)dxd\alpha^{\prime}d\alpha^{\prime\prime}
\label{eq:01-23}%
\end{equation}
The aim is to express Eq. \eqref{eq:01-23} in integrand factorized form as per
Eq.\ \eqref{eq:67-4}. We insert the following three expressions in the right
hand side of Eq.\ \eqref{eq:01-23}.%
\begin{equation}
\delta(\alpha^{\prime\prime}-x)=\int T(\beta^{\prime},\alpha^{\prime\prime
})T^{\dagger}(x,\beta^{\prime})d\beta^{\prime} \label{eq25}%
\end{equation}%
\begin{equation}
T^{\dagger}(\alpha^{\prime},\beta^{\prime})\frac{1}{T^{\dagger}(\alpha
^{\prime},\beta^{\prime})}=1 \label{eq:77-25}%
\end{equation}%
\begin{equation}
\int\delta(\alpha^{\prime}-\bar{\alpha})\delta(\beta^{\prime}-\bar{\beta
})d\bar{\alpha}d\bar{\beta}=1 \label{eq27}%
\end{equation}
to obtain
\begin{align}
\left\langle \mathbf{g}\right\rangle  &  =\int\mathcal{A}^{\ast}%
(\alpha^{\prime})\mathcal{A}(\alpha^{\prime\prime})\,\ T(\beta^{\prime}%
,\alpha^{\prime\prime})T^{\dagger}(\alpha^{\prime},\beta^{\prime})\frac
{1}{T^{\dagger}(\bar{\alpha},\bar{\beta})}\delta(\alpha^{\prime}-\bar{\alpha
})\delta(\beta^{\prime}-\bar{\beta})\nonumber\\
&  \delta(\bar{\alpha}-x)\mathbf{g}T^{\dagger}(x,\bar{\beta})dxd\alpha
^{\prime}d\alpha^{\prime\prime}d\bar{\alpha}d\bar{\beta}d\beta^{\prime}
\label{eq:01-28}%
\end{align}
We complete the integrand factorization by separating the delta functions into
a product of two terms, one depending only on $\alpha^{\prime}$ and
$\beta^{\prime},$ and the other only on $\bar{\alpha}$ and $\bar{\beta},$
\begin{equation}
\delta(\alpha^{\prime}-\bar{\alpha})\delta(\beta^{\prime}-\bar{\beta})=\left(
\frac{1}{\pi\hslash}\right)  ^{2}%
{\displaystyle\iint}
e^{-2i(\alpha-\alpha^{\prime})(\beta-\beta^{\prime})/\hslash}e^{2i(\alpha
-\bar{\alpha})(\beta-\bar{\beta})/\hslash}d\alpha d\beta\label{01-29}%
\end{equation}
In addition, this is how the $\alpha$ and $\beta$ of Eq. \eqref{eq:67-4} are introduced.

Inserting Eq.\ \eqref{01-29} into Eq.\ \eqref{eq:01-28}, we obtain%
\begin{align}
\left\langle \mathbf{g}\right\rangle  &  =\left(  \frac{1}{\pi\hslash}\right)
^{2}\int\mathcal{A}^{\ast}(\alpha^{\prime})\mathcal{A}(\alpha^{\prime\prime
})\,\ T(\beta^{\prime},\alpha^{\prime\prime})T^{\dagger}(\alpha^{\prime}%
,\beta^{\prime})e^{-2i(\alpha-\alpha^{\prime})(\beta-\beta^{\prime})/\hslash
}\frac{1}{T^{\dagger}(\bar{\alpha},\bar{\beta})}e^{2i(\alpha-\bar{\alpha
})(\beta-\bar{\beta})/\hslash}\nonumber\\
&  \delta(\bar{\alpha}-x)\mathbf{g}T^{\dagger}(x,\bar{\beta})dxd\alpha
^{\prime}d\alpha^{\prime\prime}d\bar{\alpha}d\bar{\beta}d\beta^{\prime}d\beta
d\alpha\label{eq:67-26}%
\end{align}
This achieves the factorization of the integrand because we can write
Eq.\ \eqref{eq:67-26} as
\begin{align}
\left\langle \mathbf{g}\right\rangle  &  =\int\left\{  \frac{1}{\pi\hslash
}\int\mathcal{A}^{\ast}(\alpha^{\prime})\mathcal{A}(\alpha^{\prime\prime
})T(\beta^{\prime},\alpha^{\prime\prime})T^{\dagger}(\alpha^{\prime}%
,\beta^{\prime})e^{-2i(\alpha-\alpha^{\prime})(\beta-\beta^{\prime})/\hslash
}d\alpha^{\prime}\,d\beta^{\prime}\,d\alpha^{\prime\prime}\right\} \nonumber\\
&  \left\{  \frac{1}{\pi\hslash}\frac{1}{T^{\dagger}(\bar{\alpha},\bar{\beta
})}e^{2i(\alpha-\bar{\alpha})(\beta-\bar{\beta})/\hslash}\delta(\bar{\alpha
}-x)\mathbf{g}T^{\dagger}(x,\bar{\beta})dxd\bar{\alpha}d\bar{\beta}\right\}
d\alpha d\beta\, \label{eq:01-33}%
\end{align}
where the first term depends only on the wave function and the second only on
the operator $\mathbf{g.}$ Therefore%
\begin{equation}
\left\langle \mathbf{g}\right\rangle =\int g_{W}(\alpha,\beta)P_{W}%
(\alpha,\beta)d\alpha d\beta\,
\end{equation}
where the quasi-probability distribution is
\begin{equation}
P_{W}(\alpha,\beta)=\,\frac{1}{\pi\hslash}\iiint\mathcal{A}^{\ast}%
(\alpha^{\prime})\mathcal{A}(\alpha^{\prime\prime})\,T^{\dagger}%
(\alpha^{\prime},\beta^{\prime})T(\beta^{\prime},\alpha^{\prime\prime
})e^{-2i(\alpha-\alpha^{\prime})(\beta-\beta^{\prime})/\hslash}\,d\alpha
^{\prime}\,d\beta^{\prime}\,d\alpha^{\prime\prime} \label{eq:01-34}%
\end{equation}
and the corresponding c-function
is given by
\begin{equation}
g_{W}(\alpha,\beta)=\frac{1}{\pi\hslash}\int e^{2i(\alpha-\bar{\alpha}%
)(\beta-\bar{\beta})/\hslash}\,\frac{1}{T^{\dagger}(\bar{\alpha},\bar{\beta}%
)}\ \delta(\bar{\alpha}-x)\mathbf{g}T^{\dagger}(x,\bar{\beta})d\bar{\beta
}d\bar{\alpha}dx \label{eq:01-35}%
\end{equation}

\noindent\textbf{Alternate forms. }In Eq.\ \eqref{eq:01-34} note that
\begin{equation}
\int\mathcal{A}(\alpha^{\prime\prime})T(\beta^{\prime},\alpha^{\prime\prime
})\ d\alpha^{\prime\prime}=\mathcal{B}(\beta^{\prime})
\end{equation}
and hence we can write
\begin{equation}
P_{W}(\alpha,\beta)=\,\frac{1}{\pi\hslash}%
{\displaystyle\iint}
\mathcal{A}^{\ast}(\alpha^{\prime})\mathcal{B}(\beta^{\prime})T^{\dagger
}(\alpha^{\prime},\beta^{\prime})e^{-2i(\alpha-\alpha^{\prime})(\beta
-\beta^{\prime})/\hslash}\,d\alpha^{\prime}\,d\beta^{\prime}\,
\end{equation}

Also, making the transformation in Eq.\ \eqref{eq:01-34}
\begin{equation}
\alpha^{\prime}=u-\hslash\tau/2\hspace{0.2in};\hspace{0.2in}\alpha
^{\prime\prime}=u+\hslash\tau/2
\end{equation}
with $d\alpha^{\prime}\,d\alpha^{\prime\prime}=\hslash dud\tau,$ we obtain
\begin{align}
P_{W}(\alpha,\beta)  &  =\,{\frac{1}{{\pi}}}\iiint\mathcal{A}^{\ast}%
(u-\hslash\tau/2)\mathcal{A}(u+\hslash\tau/2)\,T^{\dagger}(u-\hslash
\tau/2,\beta^{\prime})T(\beta^{\prime},u+\hslash\tau/2)\nonumber\\
&  e^{-i2(\alpha-u)(\beta-\beta^{\prime})/\hslash}\,e^{-i(\beta-\beta^{\prime
})\tau}\,d\beta^{\prime}dud\tau\label{eq:01-41}%
\end{align}

\bigskip

\noindent\textbf{Marginals.} To show that the marginals are satisfied, we
integrate Eq.\ \eqref{eq:01-34} with respect to $\beta,$
\begin{equation}
\int P_{W}(\alpha,\beta)d\beta=\,%
{\displaystyle\iint}
\mathcal{A}^{\ast}(\alpha)\mathcal{A}(\alpha^{\prime\prime})\,T^{\dagger
}(\alpha,\beta^{\prime})T(\beta^{\prime},\alpha^{\prime\prime})\,d\alpha
^{\prime\prime}d\beta^{\prime}%
\end{equation}
Using Eq.\ \eqref{eq19}\ we obtain%
\begin{equation}
\int P_{W}(\alpha,\beta)d\beta=\left\vert \mathcal{A}(\alpha)\right\vert ^{2}%
\end{equation}
which is the marginal for the $\alpha$ variable. To obtain the $\beta$
marginal we integrate $P_{W}(\alpha,\beta)$ with respect to $\alpha$ obtain
\begin{align}
\int P_{W}(\alpha,\beta)d\alpha &  =%
{\displaystyle\iint}
\mathcal{A}^{\ast}(\alpha^{\prime})\mathcal{A}(\alpha^{\prime\prime}%
)\,T^{\ast}(\beta,\alpha^{\prime})T(\beta,\alpha^{\prime\prime})\,d\alpha
^{\prime}\,\,d\alpha^{\prime\prime}\\
&  =\left\vert \int\mathcal{A}(\alpha)T(\beta,\alpha)\ d\alpha\right\vert
^{2}=\left\vert \mathcal{B}(\beta)\right\vert ^{2}%
\end{align}
which is the marginal for the $\beta$ variable.

\bigskip

\noindent\textbf{Position-momentum case. }To specialize to the $qp$ case, we
take
\begin{align}
\alpha &  \rightarrow q=\text{position}\label{eq:67-40a}\\
\beta &  \rightarrow p=\text{momentum}\label{eq:67-40b}\\
\mathcal{A}(\alpha)  &  \rightarrow\psi(q) \label{eq:67-40c}%
\end{align}
in which case the transformation matrix is
\begin{equation}
T(p,q)=\frac{1}{\sqrt{2\pi\hslash}}e^{-iqp/\hslash} \label{eq:67-41}%
\end{equation}
Inserting these into Eq.\ \eqref{eq:01-41}, we obtain that
\begin{equation}
P_{W}(\alpha,\beta)\,=\frac{1}{2\pi}\,\int\psi^{\ast}(q-\hslash\tau
/2)\psi(q+\hslash\tau/2)e^{-ip\tau}d\tau\,\,\,\,
\end{equation}
which is the Wigner distribution. This is the reason we have called
Eq.\ \eqref{eq:01-41} Wigner type for arbitrary operators.

For the correspondence rule, inserting
Eqs.\ \eqref{eq:67-40a}--\eqref{eq:67-41} into Eq.\ \eqref{eq:01-35}, one
obtains that
\begin{equation}
g_{W}(q,p)=\hslash%
{\displaystyle\iint}
e^{i\tau p}\,\ \delta(q-\hslash\tau/2-x)\mathbf{g(x,p}_{x}\mathbf{)}%
\delta(q+\hslash\tau/2-x)dx\ d\tau\label{eq:01-53}%
\end{equation}
It has been previously shown \cite{kim} that Eq.\ \eqref{eq:01-53} is the
inverse Weyl, that is, it is the c-function obtained from the quantum operator.

\section{Dirac notation derivation}

It is of interest to give a derivation of Eqs.\ \eqref{eq:01-34} and
\eqref{eq:01-35} in terms of Dirac notation and method. Starting with
\begin{equation}
\text{Tr}\left\{  \boldsymbol{\mathbf{g}\rho}\right\}  =\iint\,\left\langle
\alpha^{\prime}|\mathbf{g|}\beta^{\prime}\right\rangle \left\langle
\beta^{\prime}|\boldsymbol{\rho|}\alpha^{\prime}\right\rangle d\alpha^{\prime
}d\beta^{\prime} \label{eq:77-50}%
\end{equation}
Inserting
\begin{equation}
\left\langle \alpha^{\prime}|\beta^{\prime}\right\rangle \frac{1}{\left\langle
\alpha^{\prime}|\beta^{\prime}\right\rangle }=1
\end{equation}
and
\begin{equation}
\delta(\alpha^{\prime}-\bar{\alpha})\delta(\beta^{\prime}-\bar{\beta}%
)=\frac{1}{(\pi\hbar)^{2}}\iint\,e^{-2i(\alpha-\alpha^{\prime})(\beta
-\beta^{\prime})/\hbar}e^{2i(\alpha-\bar{\alpha})(\beta-\bar{\beta})/\hslash
}d\alpha d\beta
\end{equation}
into Eq.\ \eqref{eq:77-50} we obtain
\begin{align}
\text{Tr}\left\{  \boldsymbol{\mathbf{g}\rho}\right\}   &  =\frac{1}{(\pi
\hbar)^{2}}\iint d\alpha d\beta\iint d\alpha^{\prime}d\beta^{\prime
}\,\left\langle \beta^{\prime}|\boldsymbol{\rho}|\alpha^{\prime}\right\rangle
\left\langle \alpha^{\prime}|\beta^{\prime}\right\rangle e^{-2i(\alpha
-\alpha^{\prime})(\beta-\beta^{\prime})/\hbar}\nonumber\\
&  \times\iint d\bar{\alpha}d\bar{\beta}\,\frac{1}{\left\langle \bar{\alpha
}|\bar{\beta}\right\rangle }\left\langle \bar{\alpha}|\mathbf{g|}{\bar{\beta}%
}\right\rangle e^{2i(\alpha-\bar{\alpha})(\beta-\bar{\beta})/\hslash}%
\end{align}
which achieves the factorization of the integrand. Therefore the phase space
distribution is
\begin{equation}
P_{W}(\alpha,\beta)=\frac{1}{\pi\hbar}\iint d\alpha^{\prime}d\beta^{\prime
}\left\langle \beta^{\prime}|\mathbf{\rho|}\alpha^{\prime}\right\rangle
\left\langle \alpha^{\prime}|\beta^{\prime}\right\rangle e^{-2i(\beta
-\beta^{\prime})(\alpha-\alpha^{\prime})/\hbar} \label{eq:06-8}%
\end{equation}
and the corresponding c-number is
\begin{equation}
g_{W}(\alpha,\beta)=\frac{1}{\pi\hbar}\iint d\bar{\alpha}d\bar{\beta}%
\,\frac{1}{\left\langle \bar{\alpha}|\bar{\beta}\right\rangle }\left\langle
\bar{\alpha}|\mathbf{g|}\bar{\beta}\right\rangle e^{2i(\beta-\bar{\beta
})(\alpha-\bar{\alpha})/\hbar} \label{eq:06-9}%
\end{equation}
Equivalence of Eqs.\ \eqref{eq:06-8} and \eqref{eq:06-9} with
Eqs.\ \eqref{eq:01-34} and \eqref{eq:01-35} is readily shown.

\section{Margenau-Hill type case}

We start with Eq.\ \eqref{eq:01-15}
\begin{equation}
\left\langle \mathbf{g}\right\rangle =%
{\displaystyle\iiint}
\mathcal{A}^{\ast}(\alpha)\mathcal{A}(\alpha^{\prime})\delta(\alpha
-x)\mathbf{g}\delta(\alpha^{\prime}-x)d\alpha d\alpha^{\prime}dx
\end{equation}
and we see that replacing
\begin{equation}
\delta(\alpha^{\prime}-x)=\int T^{\dagger}(x,\beta)T(\beta,\alpha^{\prime
})d\beta
\end{equation}
and inserting Eq.\ \eqref{eq:77-25} immediately achieves the factorization of
the integrand,
\begin{align}
\left\langle \mathbf{g}\right\rangle  &  =\int\left\{  \mathcal{A}^{\ast
}(\alpha)\,T^{\dagger}(\alpha,\beta)\int\mathcal{A}(\alpha^{\prime}%
)T(\beta,\alpha^{\prime})\,d\alpha^{\prime}\right\}  \nonumber\\
&  \left\{  \left(  T^{\dagger}(\alpha,\beta)\right)  ^{-1}\int\delta
(\alpha-x)\mathbf{g}T^{\dagger}(x,\beta)\ dx\right\}  d\beta\ d\alpha
\end{align}
Therefore%
\begin{equation}
\left\langle \mathbf{g}\right\rangle =\int g(\alpha,\beta)P(\alpha
,\beta)\ d\alpha d\beta
\end{equation}
with%
\begin{align}
P_{\text{MH}}(\alpha,\beta) &  =\mathcal{A}^{\ast}(\alpha)\,T^{\dagger}%
(\alpha,\beta)\int\mathcal{A}(\alpha^{\prime})T(\beta,\alpha^{\prime
})\,d\alpha^{\prime}\\
g_{\text{MH}}(\alpha,\beta) &  =\left(  T^{\dagger}(\alpha,\beta)\right)
^{-1}\int\delta(\alpha-x)\mathbf{g}T^{\dagger}(x,\beta)\ dx
\end{align}
\noindent\textbf{Alternate forms. }Also, since
\begin{equation}
\mathcal{B}(\beta)=\int\mathcal{A}(\alpha^{\prime})T(\beta,\alpha^{\prime
})\ d\alpha^{\prime}%
\end{equation}
we may also write%
\begin{align}
P_{MH}(\alpha,\beta) &  =\mathcal{A}^{\ast}(\alpha)\,T^{\dagger}(\alpha
,\beta)\mathcal{B}(\beta)\,\label{eq69}\\
g_{MH}(\alpha,\beta) &  =\left(  T^{\dagger}(\alpha,\beta)\right)  ^{-1}%
\int\delta(\alpha-x)\mathbf{g}T^{\dagger}(x,\beta)\ dx\label{eq70}%
\end{align}
We note that $P_{MH}^{\ast}(\alpha,\beta)$ is also a quasi-distribution.

\bigskip

\noindent\textbf{Position-momentum case}. To specialize to the $qp$ case, as
per Eqs.\ \eqref{eq:67-40a}--\eqref{eq:67-41} we have%
\begin{equation}
\mathcal{B}(p)=\frac{1}{\sqrt{2\pi\hslash}}\int\psi(q^{\prime})e^{-iq^{\prime
}p/\hslash}\ dq^{\prime}=\varphi(p)
\end{equation}
which is the momentum wave function. Hence, substituting into
Eq.\ \eqref{eq69} we obtain
\begin{equation}
P_{MH}(q,p)=\frac{1}{\sqrt{2\pi\hslash}}\psi^{\ast}(q)e^{iqp/\hslash}%
\varphi(p)
\end{equation}
which is a distribution first mentioned by Kirkwood \cite{kirk} and derived by
Margenau and Hill \cite{mh} and studied by Mehta \cite{mehta} and others, and
is known in the field of time-frequency analysis as the Rihaczek distribution
\cite{ri}.

Using Eq.\ \eqref{eq70}, we obtain that
\begin{equation}
g_{MH}(q,p)=e^{-iqp/\hslash}\mathbf{g\left(  q,p\right)  }e^{iqp/\hslash}\
\end{equation}

\section{General class}

Cohen obtained the general class of quasi-distributions for the $qp$ case by
introducing a kernel function that characterizes each quasi-distribution and
its correspondence rule \cite{cohen66}. By taking all possible kernel
functions, one obtains the totality of bilinear quasi-distributions and rules
of association \cite{lee,cw,jeong,zam}. We now show how to generate an
infinite number of quasi-distributions for arbitrary operators. Again, we
start with the quantum expectation value
\begin{equation}
\left\langle \mathbf{g}\right\rangle =\int\mathcal{A}^{\ast}(\alpha^{\prime
})\mathcal{A}(\alpha^{\prime\prime})\delta(\alpha^{\prime}-x)\mathbf{g}%
\delta(\alpha^{\prime\prime}-x)dxd\alpha^{\prime}d\alpha^{\prime\prime}%
\end{equation}
Inserting the three expressions Eqs. \eqref{eq25}-\eqref{eq27}, which we
repeat here
\begin{equation}
\delta(\alpha^{\prime\prime}-x)=\int T(\beta^{\prime},\alpha^{\prime\prime
})T^{\dagger}(x,\beta^{\prime})d\beta^{\prime}%
\end{equation}%
\begin{equation}
T^{\dagger}(\alpha^{\prime},\beta^{\prime})\frac{1}{T^{\dagger}(\alpha
^{\prime},\beta^{\prime})}=1
\end{equation}%
\begin{equation}
\int\delta(\alpha^{\prime}-\bar{\alpha})\delta(\beta^{\prime}-\bar{\beta
})d\bar{\alpha}d\bar{\beta}%
\end{equation}
we obtain
\begin{align}
\left\langle \mathbf{g}\right\rangle  &  =\int\mathcal{A}^{\ast}%
(\alpha^{\prime})\mathcal{A}(\alpha^{\prime\prime})T(\beta^{\prime}%
,\alpha^{\prime\prime})T^{\dagger}(\alpha^{\prime},\beta^{\prime})\frac
{1}{T^{\dagger}(\bar{\alpha},\bar{\beta})}\delta(\alpha^{\prime}-\bar{\alpha
})\delta(\beta^{\prime}-\bar{\beta})\nonumber\\
&  \delta(\bar{\alpha}-x)\mathbf{g}T^{\dagger}(x,\bar{\beta})dxd\alpha
^{\prime}d\alpha^{\prime\prime}d\bar{\alpha}d\bar{\beta}d\beta^{\prime}%
\end{align}
To introduce the phase space variables $\alpha$ and $\beta,$ we insert%
\begin{equation}
\delta(\bar{\alpha}-\alpha^{\prime})\delta(\bar{\beta}-\beta^{\prime}%
)=\frac{1}{(2\pi)^{4}}\int\frac{\Phi(\theta,\tau)e^{i\theta(\alpha^{\prime
}-\alpha)}e^{i\tau\theta\hslash/2}e^{i\tau(\beta^{\prime}-\beta)}}{\Phi
(\theta^{\prime},\tau^{\prime})e^{i\theta^{\prime}(\bar{\alpha}-\alpha
)}e^{i\tau^{\prime}\theta^{\prime}\hslash/2}e^{i\tau^{\prime}(\bar{\beta
}-\beta)}}\,\,d\theta d\tau d\theta^{\prime}d\tau^{\prime}d\alpha
d\beta\label{eq:01-81}%
\end{equation}
where $\Phi(\theta,\tau)$ is the kernel function that characterizes the
distribution. We obtain%

\begin{align}
\left\langle \mathbf{g}\right\rangle  &  =\int\left\{  \frac{1}{4{\pi}^{2}%
}\int\mathcal{A}^{\ast}(\alpha^{\prime})\mathcal{A}(\alpha^{\prime\prime
})T(\beta^{\prime},\alpha^{\prime\prime})T^{\dagger}(\alpha^{\prime}%
,\beta^{\prime})\Phi(\theta,\tau)e^{i\theta(\alpha^{\prime}-\alpha)}%
e^{i\tau\theta\hslash/2}e^{i\tau(\beta^{\prime}-\beta)}d\alpha^{\prime
}\,d\beta^{\prime}\,d\alpha^{\prime\prime}d\theta d\tau\,\right\} \nonumber\\
&  \left\{  \int\frac{1}{4{\pi}^{2}}\frac{1}{\Phi(\theta^{\prime},\tau
^{\prime})e^{i\theta^{\prime}(\bar{\alpha}-\alpha)}e^{i\tau^{\prime}%
\theta^{\prime}\hslash/2}e^{i\tau^{\prime}(\bar{\beta}-\beta)}}\frac
{1}{T^{\dagger}(\bar{\alpha},\bar{\beta})}\delta(\bar{\alpha}-x)\mathbf{g}%
T^{\dagger}(x,\bar{\beta})dxd\bar{\alpha}d\bar{\beta}d\theta^{\prime}%
d\tau^{\prime}\right\}  d\alpha d\beta
\end{align}
which achieves the factorization of the integrand. Therefore%
\begin{equation}
\left\langle \mathbf{g}\right\rangle =\int g_{\Phi}(\alpha,\beta)P_{\Phi
}(\alpha,\beta)d\alpha d\beta
\end{equation}
with the quasi-distribution given by%
\begin{equation}
P_{\Phi}(\alpha,\beta)=\frac{1}{4{\pi}^{2}}\,\int\Phi(\theta,\tau
)\mathcal{A}^{\ast}(\alpha^{\prime})\mathcal{A}(\alpha^{\prime\prime
})\,T(\beta^{\prime},\alpha^{\prime\prime})T^{\dagger}(\alpha^{\prime}%
,\beta^{\prime})e^{i\theta(\alpha^{\prime}-\alpha)}e^{i\tau\theta\hslash
/2}e^{i\tau(\beta^{\prime}-\beta)}d\alpha^{\prime}\,d\beta^{\prime}%
\,d\alpha^{\prime\prime}\,\,d\theta d\tau\label{eq:02-66b}%
\end{equation}
and the corresponding c-funtion is given by%
\begin{equation}
g_{\Phi}(\alpha,\beta)=\frac{1}{4{\pi}^{2}}\int\frac{e^{-i\theta^{\prime}%
(\bar{\alpha}-\alpha)}e^{-i\tau^{\prime}\theta^{\prime}\hslash/2}%
e^{-i\tau^{\prime}(\bar{\beta}-\beta)}}{T^{\dagger}(\bar{\alpha},\bar{\beta
})\Phi(\theta^{\prime},\tau^{\prime})}\delta(\bar{\alpha}-x)\mathbf{g}%
T^{\dagger}(x,\bar{\beta})dxd\bar{\alpha}d\bar{\beta}d\theta^{\prime}%
d\tau^{\prime}%
\end{equation}

\bigskip

If we take $\Phi(\theta,\tau)=1$, then we obtain the Wigner-type case above,
and if one takes $\Phi(\theta,\tau)=$ $e^{i\theta\tau\hslash/2}$ we obtain Eq.\ \eqref{eq69}.

\bigskip

\noindent\textbf{Alternate form. }Making the transformation in
Eq.\ \eqref{eq:02-66b}%
\begin{equation}
\alpha^{\prime}=u-\hslash\tau/2\hspace{0.2in};\hspace{0.2in}\alpha
^{\prime\prime}=u+\hslash\tau/2
\end{equation}
One obtains
\begin{align}
P_{\Phi}(\alpha,\beta)  &  =\frac{\hslash}{4{\pi}^{2}}\,\int\Phi(\theta
,\tau^{\prime})\mathcal{A}^{\ast}(u-\hslash\tau/2)\mathcal{A}(u+\hslash
\tau/2)\,T^{\dagger}(u-\hslash\tau/2,\beta^{\prime})T(\beta^{\prime}%
,u+\hslash\tau/2)\nonumber\\
&  \,e^{i\theta\lbrack u-\alpha-\hslash(\tau-\tau^{\prime})/2]}e^{i\tau
^{\prime}(\beta^{\prime}-\beta)}d\beta^{\prime}d\theta d\tau^{\prime}%
dud\tau\label{eq:01-89}%
\end{align}

\noindent\textbf{Position-momentum case. }To specialize to the position
momentum case we use Eqs.\ \eqref{eq:67-40a}--\eqref{eq:67-41}; inserting them
into Eq.\ \eqref{eq:01-89} we obtain
\begin{equation}
P_{\Phi}(q,p)\,=\frac{1}{4{\pi}^{2}}%
{\displaystyle\iiint}
\Phi(\theta,\tau)\psi^{\ast}(u-\tau\hslash/2)\psi(u+\tau\hslash/2)\,e^{i\theta
(u-q)-i\tau p}du\,d\theta d\tau\label{eq:01-91}%
\end{equation}
which is the generalized quasi-probability distribution obtained by Cohen
\cite{cohen66}. All bilinear position-momentum distributions are generated by
Eq.\ \eqref{eq:01-91} by taking different functions for $\Phi(\theta,\tau).$
Similarly all bilinear distributions for arbitrary operators are generated by Eq.\ \eqref{eq:01-89}.

For the correspondence rule we have
\begin{equation}
g_{\Phi}(q,p)=\frac{1}{(2\pi)^{2}}\int\frac{1}{\Phi(\theta,\tau)}\,e^{i\tau
p}e^{-i\theta\bar{\alpha}}\delta(\bar{\alpha}+q-\hbar\tau/2-x)\mathbf{g(x,p}%
_{x})\delta(\bar{\alpha}+q+\hbar\tau/2-x)d\bar{\alpha}dxd\theta d\tau\label{a}%
\end{equation}
In the original formulation for generating all correspondence rules, the
operator is given by \cite{cohen66}%

\begin{equation}
\mathbf{g_{\Phi}(x,p}_{x})\,=%
{\displaystyle\iint}
\widehat{g}_{\Phi}(\theta,\tau)\Phi(\theta,\tau)\,e^{i\theta\mathbf{x+}%
i\tau\mathbf{p}_{x}}\,\,d\theta\,d\tau\label{c}%
\end{equation}
That Eq. \eqref{a} obtains the classical function appearing in Eq.\ \eqref{c}
in terms of the operator can be readily verified by substituting
Eq.\ \eqref{c} into Eq.\ \eqref{a}

\noindent\textbf{Commuting operators. }Commuting operators have common
eigenfunctions and we may write
\begin{align}
\mathbf{a}u_{\alpha}(q)\,  &  =\,\alpha\ u_{\alpha}(q)\\
\mathbf{b}u_{\alpha}(q)\,\,  &  =\gamma(\alpha)\ u_{\alpha}(q)
\end{align}
Cohen has shown, using the characteristic function operator method, that for
all distributions
\begin{equation}
P(\alpha,\beta)=\delta(\beta-\gamma(\alpha))\left\vert \mathcal{A}%
(\alpha)\right\vert ^{2}%
\end{equation}
This results may also be derived from Eq.\ \eqref{eq:02-66b}.

\section{Characteristic function}

For two random variables, $\alpha$ and $\beta,$ the characteristic function,
$M(\theta,\tau)$, and distribution, $P(\alpha,\beta)$, are respectively
\begin{equation}
M(\theta,\tau)=\langle e^{i\theta\alpha+i\tau\beta}\rangle=\iint
e^{i\theta\alpha+i\tau\beta}P(\alpha,\beta)d\alpha d\beta\
\end{equation}%
\begin{equation}
P(\alpha,\beta)={\frac{1}{{4\pi^{2}}}}\iint e^{-i\theta\alpha-i\tau\beta
}M(\theta,\tau)d\alpha d\beta
\end{equation}
Consider now the characteristic function for the Wigner type,
Eq.\ \eqref{eq:01-34}, and of the general class, Eq.\ \eqref{eq:02-66b},
\begin{equation}
M_{W}(\theta,\tau)=\,\frac{1}{\pi\hslash}\int e^{i\theta\alpha+i\tau\beta
}\mathcal{A}^{\ast}(\alpha^{\prime})\mathcal{A}(\alpha^{\prime\prime
})\,T^{\dagger}(\alpha^{\prime},\beta^{\prime})T(\beta^{\prime},\alpha
^{\prime\prime})e^{-2i(\beta-\beta^{\prime})(\alpha-\alpha^{\prime})/\hslash
}\,d\alpha^{\prime}\,d\beta^{\prime}\,d\alpha^{\prime\prime}d\alpha d\beta
\end{equation}
and%
\begin{align}
M_{\Phi}(\theta,\tau)  &  =\frac{1}{4{\pi}^{2}}\int\Phi(\theta^{\prime}%
,\tau^{\prime})\mathcal{A}^{\ast}(\alpha^{\prime})\mathcal{A}(\alpha
^{\prime\prime})\,T^{\dagger}(\alpha^{\prime},\beta^{\prime})T(\beta^{\prime
},\alpha^{\prime\prime})\nonumber\\
&  e^{i\theta^{\prime}(\alpha^{\prime}-\alpha)}e^{i\tau^{\prime}\theta
^{\prime}\hslash/2}e^{i\tau^{\prime}(\beta^{\prime}-\beta)}e^{i\theta
\alpha+i\tau\beta}d\alpha^{\prime}\,d\beta^{\prime}\,d\alpha^{\prime\prime
}d\theta^{\prime}d\tau^{\prime}d\alpha d\beta
\end{align}
Straight-forward manipulation leads to the fact that%
\begin{equation}
M_{\Phi}(\theta,\tau)=\Phi(\theta,\tau)M_{W}(\theta,\tau)
\end{equation}
This is identical in form to the $qp$ case \cite{cohen66,lee}.

\section{Example}

We consider the Hamiltonian of a particle acted on by constant force, $f,$
\begin{equation}
\mathbf{H}=f\mathbf{q}+\frac{\mathbf{p}^{2}}{2m} \label{eq:01-99}%
\end{equation}
and we seek a joint quasi-distribution of position and the Hamiltonian. We take%

\begin{equation}
\mathbf{a}=\mathbf{q},\quad\mathbf{b}=\mathbf{H}%
\end{equation}
The eigenvalue problem for $\mathbf{H}$ can be readily solved in the momentum
representation
\begin{equation}
\mathbf{b}\ \widehat{v}_{\beta}(p)\,=\left\{  \ f\mathbf{q}+\frac
{\mathbf{p}^{2}}{2m}\right\}  \widehat{v}_{\beta}(p)\,=\,\beta\widehat
{v}_{\beta}(p)
\end{equation}
or
\begin{equation}
\left\{  \ i\hslash f\frac{d}{dp}+\frac{p^{2}}{2m}\ \right\}  \widehat
{v}_{\beta}(p)\ \,=\,\beta\widehat{v}_{\beta}(p)
\end{equation}
The delta function normalized solution is
\begin{equation}
\ \widehat{v}_{\beta}(p)\ =\frac{1}{\sqrt{2\pi\hslash f}}e^{-i(\beta
p-p^{3}/6m)/\hslash f}%
\end{equation}
In the spatial domain
\begin{equation}
\mathbf{b}v_{\beta}(q)=\beta v_{\beta}(q)
\end{equation}
the eigenfunctions are
\begin{equation}
\ v_{\beta}(q)\ =\frac{1}{2\pi\hslash}\frac{1}{\sqrt{f}}\int e^{-i(\beta
p-p^{3}/6m)/\hslash f}e^{iqp/\hslash}dp
\end{equation}
and the transformation matrix,
\begin{equation}
T(\beta,\alpha)=\int v_{\beta}^{\ast}(q)u_{\alpha}(q)\,dq
\end{equation}
is
\begin{align}
T(\beta,\alpha)  &  =\frac{1}{2\pi\hslash}\frac{1}{\sqrt{f}}\iint e^{i(\beta
p-p^{3}/6m)/\hslash f}e^{-iqp/\hslash}\delta(q-\alpha)\,dqdp\\
&  =\frac{1}{2\pi\hslash}\frac{1}{\sqrt{f}}\int e^{-ip^{3}/6m\hslash
f}e^{i(\beta/f-\alpha)p/\hslash}dp
\end{align}

Since $\mathcal{A}$ is now the position wave function, $\psi,$ we can write%
\begin{equation}
P(\alpha,\beta)=\,\frac{1}{\pi\hslash}\iiint\psi^{\ast}(\alpha^{\prime}%
)\psi(\alpha^{\prime\prime})\,T^{\dagger}(\alpha^{\prime},\beta^{\prime
})T(\beta^{\prime},\alpha^{\prime\prime})e^{-2i(q-\alpha^{\prime}(\beta
-\beta^{\prime}))/\hslash}\,d\alpha^{\prime}\,d\beta^{\prime}\,d\alpha
^{\prime\prime}%
\end{equation}
with
\begin{equation}
\,T^{\dagger}(q^{\prime},\beta^{\prime})T(\beta^{\prime},q^{\prime\prime
})=\frac{1}{(2\pi\hslash)^{2}}\frac{1}{f}\iint dpdp^{\prime}e^{i(p^{3}%
-p^{\prime3})/6m\hbar f}e^{-i\beta^{\prime\prime}/\hslash f}e^{i(\alpha
^{\prime}p-\alpha^{\prime\prime}p^{\prime})/\hslash}%
\end{equation}

This could be simplified further by using the wave function in the
$\mathbf{b}$ basis, $\mathcal{B}(\beta),$
\begin{equation}
P(\alpha,\beta)=\frac{1}{(2\pi\hslash)^{2}}\frac{1}{\pi}%
{\displaystyle\iint}
\psi^{\ast}(\alpha^{\prime})\mathcal{B}(\beta^{\prime})\,T^{\dagger}%
(\alpha^{\prime},\beta^{\prime})e^{-2i(\alpha-\alpha^{\prime})(\beta
-\beta^{\prime})/\hslash}\,d\alpha^{\prime}\,d\beta^{\prime} \label{eq:03-5}%
\end{equation}
where
\begin{equation}
\mathcal{B}(\beta)=\int T(\beta,\alpha)\psi(\alpha)d\alpha
\end{equation}

\noindent\textbf{Momentum and energy. }We take
\begin{equation}
\mathbf{a}=\mathbf{p},\quad\mathbf{b}=\mathbf{H}%
\end{equation}
The eigenfunctions for momentum are then,%
\begin{equation}
u_{\alpha}(q)=\frac{1}{\sqrt{2\pi\hslash}}e^{i\alpha q/\hslash}%
\end{equation}
and the transformation matrix is
\begin{align}
T(\beta,\alpha)  &  =\int v_{\beta}^{\ast}(q)u_{\alpha}(q)\,dq\\
&  =\frac{1}{2\pi\hslash}\frac{1}{\sqrt{f}}e^{i(\beta\alpha-\alpha
^{3}/6m)/\hslash f}%
\end{align}
The quasi-distribution of momentum and energy is
\begin{equation}
P(\alpha,\beta)=\frac{1}{(2\pi\hslash)^{2}}\frac{1}{\pi}\iiint\varphi^{\ast
}(\alpha^{\prime})\mathcal{B}(\beta^{\prime})\,T^{\dagger}(\alpha^{\prime
},\beta^{\prime})e^{-2i(\alpha-\alpha^{\prime})(\beta-\beta^{\prime})/\hslash
}\,d\alpha^{\prime}\,d\beta^{\prime}%
\end{equation}
where $\varphi$ is the momentum wave function and
\begin{align}
\mathcal{A}(\alpha)  &  =\varphi(\alpha)=\frac{1}{\sqrt{2\pi\hslash}}\int
e^{-iq\alpha/\hbar}\psi(q)dq\\
\mathcal{B}(\beta)  &  =\int T(\beta,\alpha)\varphi(\alpha)d\alpha
\end{align}

\section{Conclusion}

We have obtained joint quasi-distributions for arbitrary operators. Starting
with the quantum mechanical expression for the expectation value of an
operator, we have expressed it as phase space integral where the integrand is
factorized into two factors: One depending only on the density matrix and
other only on the operator. Simultaneously with the derivation of the
quasi-distribution, one obtains the generalization of the concept of
correspondence rule for arbitrary operators. An advantage of our approach is
that it shows straightforwardly the coupling of the quasi-distribution with
its correspondence rule. Although we have presented our results for the pure
case, generalization to the density matrix follows straightforwardly.

In the classic paper by Moyal, he derived the Wigner distribution by defining
the characteristic function by way of
\begin{equation}
M(\theta,\tau)=\int\ \psi^{\ast}(q)e^{i\theta\mathbf{q}+i\tau\mathbf{p}}%
\ \psi(q)\ dq
\end{equation}
where $\mathbf{q}$ and $\mathbf{p}$ are the position and momentum operators.
For the case of position and momentum, the operator $e^{i\theta\mathbf{q}%
+i\tau\mathbf{p}}$ can be simplified and $M(\theta,\tau)$ calculated explicitly.

Scully and Cohen \cite{scully-cohen,arb,cohenieee} generalized the method by
defining the characteristic function for two arbitrary operators%
\begin{equation}
M(\theta,\tau)=\int\ \psi^{\ast}(q)e^{i\theta\mathbf{a}+i\tau\mathbf{b}}%
\psi(q)\ dq
\end{equation}
and this has been carried out for a number of cases where $e^{i\theta
\mathbf{a}+i\tau\mathbf{b}}$ can be simplified. However, the simplification of
$e^{i\theta\mathbf{a}+i\tau\mathbf{b}}$ is generally difficult
\cite{wilcox,scully-cohen,arb,cohenieee}. Of course there is a relation
between the two approaches and this will be discussed in a future paper.

\section{Acknowledgements}

The work of JSB was supported by the Robert A.\ Welch Foundation (Grant
No.\ A-1261), the Office of Naval Research (Award No.\ N00014-16-1-3054), the
Air Force Office of Scientific Research (FA9550-18-1-0141), and the King
Abdulaziz City for Science and Technology (KACST) grant.

\bigskip
\end{document}